# Wafer Scale Homogeneous Bilayer Graphene Films by Chemical Vapor Deposition

*Seunghyun Lee[§], Kyunghoon Lee[§], Zhaohui Zhong [*]*

Department of Electrical Engineering and Computer Science, University of Michigan

Ann Arbor, MI 48109, USA

[§] These authors contributed equally to this work.

*Corresponding author. Electronic mail: zzhong@umich.edu

**ABSTRACT**

The discovery of electric field induced bandgap opening in bilayer graphene opens new door for making semiconducting graphene without aggressive size scaling or using expensive substrates. However, bilayer graphene samples have been limited to $\mu m^2$ size scale thus far, and synthesis of wafer scale bilayer graphene posts tremendous challenge. Here we report homogeneous bilayer graphene films over at least 2 inch × 2 inch area, synthesized by chemical vapor deposition on copper foil and subsequently transferred to arbitrary substrates. The bilayer nature of graphene film is verified by Raman spectroscopy, atomic force microscopy (AFM), and transmission electron microscopy (TEM). Importantly, spatially resolved Raman spectroscopy confirms a bilayer coverage of over 99%. The homogeneity of the film is further supported by electrical transport measurements on dual-gate bilayer graphene transistors, in which bandgap opening is observed in 98% of the devices.

**KEYWORDS**

Graphene, bilayer, chemical vapor deposition, wafer scale, bandgap opening

Single and few-layer graphene[1-5] are promising materials for post-silicon electronics because of their potential of integrating bottom-up nanomaterial synthesis with top-down lithographic fabrication at wafer scale.[4,6] However, single layer graphene is intrinsically semimetal; introducing energy bandgap requires patterning nanometer-width graphene ribbons[7-9] or utilizing special substrates.[10-12] Bilayer graphene, instead, has an electric field induced bandgap up to 250 meV,[13-18] thus eliminating the need for extreme scaling or costly substrates. Furthermore, exciton binding energies in bilayer graphene are also found to be tunable with electric field.[19] The unique ability of controlling the bandgap and the exciton energy can lead to new possibilities of bilayer graphene based electronics and photonics.

To date, most bilayer graphene samples are fabricated using mechanical exfoliation of graphite[15-18], which have limited sizes of $\mu m^2$ and are certainly not scalable. Recent developments in CVD method have allowed successful production of large scale single-layer graphene on metal substrate.[20-24] However, the synthesis of uniform bilayer graphene film remains extremely challenging. Here we report the first synthesis of wafer scale bilayer graphene film over at least 2 inch × 2 inch area, limited only by our synthesis apparatus. Our method is based on CVD growth of bilayer graphene on copper surface, and is characterized by the depletion of hydrogen, high vacuum, and most importantly, slower cooling rate compared to previous single-layer graphene synthesis.[21,22,24] The optimal bilayer graphene film is grown at 1000 $^o$C for 15 minutes, with growth pressure of 0.5 Torr, $CH_4$ flow rate of 70 sccm, and a cooling rate of 18°C/min (0.3 °C/s) (Supporting Information Fig. S1).

Figure 1a shows photographic image of a wafer scale (2 inch × 2 inch) bilayer graphene film transferred onto a 4 inch silicon wafer with 280 nm thick $SiO_2$. A typical optical microscope image (Fig. 1b) of the transferred bilayer graphene film shows almost no color variation except for the region where the film is torn and folded (lower right of Fig. 1b). To identify the number of layers for our graphene sample, the film thickness is first measured using AFM (Fig. 1c). Height profiles across patterned graphene edges show that thickness of our graphene samples range from 0.9 nm to 1.3 nm, suggesting number of graphene layers below 3.[23]

We further performed Raman spectroscopy measurements (Renishaw spectrometer at 514nm) on ten randomly chosen spots across the film, and compared them with reference samples prepared by mechanically exfoliating Kish graphite.[17,25,26] The red curve in Fig. 1c represents a typical Raman spectrum from our sample. Two peaks are clearly visible between Raman shift of 1250 cm$^{-1}$ – 2850 cm$^{-1}$, corresponding to the G band (~1595 cm$^{-1}$) and 2D band (~2691 cm$^{-1}$), respectively.[26-30] Importantly, the spectrum exhibits several distinctive features. First, 2D band shows higher peak intensity than G band with the 2D-to-G intensity ratio $I_{2D}/I_G$ ~2.31, suggesting the number of graphene layers less than 3.[24,27,29] Second, the full width at half maximum (FWHM) of 2D band peak is measured to be ~45 cm$^{-1}$, exceeding the cut-off of ~30 cm$^{-1}$ for single-layer graphene.[25,26,28] Third, the 2D band peak cannot be fitted with single Lorentzian (Supporting information Fig. S2a), but fitting from four Lorentzian peaks with a FWHM of 24 cm$^{-1}$ yields excellent agreement (Supporting information Fig. S2b).[26-28] Raman spectra taken from the other 9 spots show similar features with the 2D band FWHM of 43~53 cm$^{-1}$. These observations are strong reminiscent of characteristic bilayer graphene Raman spectrum. In addition, reference Raman spectra taken under identical conditions from exfoliated single-layer (green curve) and bilayer (blue curve) graphene are also presented in Fig. 1c. Exfoliated single-layer graphene shows a 2D band FWHM of 24 cm$^{-1}$ and $I_{2D}/I_G$ of 3.79, while exfoliated bilayer graphene shows a FWHM of 46 cm$^{-1}$ and $I_{2D}/I_G$ of 2.25. Together, the AFM height measurements, the Raman spectra, and the direct comparison with the exfoliated samplesclearly supports the bilayer nature of our CVD synthesized graphene film. We also measured the D band to G band intensity ratio, $I_D/I_G$, of our bilayer graphene sample to be around 0.11~0.3, indicating a relatively low defect density.

TEM selected area electron diffraction pattern was measured to further characterize the graphene film (Figure 2a). The six-fold symmetry is clearly visible and Bravais-Miller (hkil) indices are used to label the diffraction peaks. Importantly, the diffraction intensities of inner peaks from equivalent planes {1100} are always higher than outer peaks from {2100}. The intensity ratios of $I_{1010}/I_{-1-120}$ and $I_{-1100}/I_{1-210}$ are close to 0.28 (Fig. 2b), indicating that the film is not a single layer and it retains AB stacking structure.[31-33] We further studied the tilt angle-dependent diffraction peak intensity for both inner and outer peaks. As shown in Fig. 2c, both (0-110) and (-1-120) peaks show strong intensity modulation with tilt angle, and both peaks can be suppressed completely at certain angle. It is known that monolayer graphene has only zero order Laue zone and weak intensity modulation is expected at any angle.[32,33] Our TEM results again agree with AFM and Raman measurements for the bilayer nature of the graphene film. We also notice additional diffraction spots, which are caused by the residues on the film due to insufficient sample cleaning.

To further evaluate the uniformity of CVD grown bilayer graphene film, we performed spatially resolved Raman spectroscopy. Here, identifications of the number of graphene layers rely on combination of the FWHM of 2D band[24,26-28,30] and peak intensity ratio $I_{2D}/I_G$.[23,24,27,29] Figure 3a shows a color map of the 2D band peak width over 30 μm by 30 μm area, with FWHM values ranging from 42 cm$^{-1}$ (dark color) to 63cm$^{-1}$ (bright yellow). The data show uniformly distributed red color with only a few localized yellow spots. The peak intensity ratios $I_{2D}/I_G$ are also mapped in color (Fig. 3b) over the same area, with values ranging from 0.37 (dark color) to 3.77 (bright yellow). Comparisons between Fig. 3a and Fig. 3b reveal that larger peak widths are consistent with smaller $I_{2D}/I_G$ ratios. Furthermore, Fig. 3c compares the Raman spectra taken from three representative spots indicated using green, pink, and blue circles. Raman spectrum taken at green-circled spot has the largest FWHM (62.9 cm$^{-1}$) and smallest $I_{2D}/I_G$ (0.37), indicating trilayer graphene (Supporting information Fig. S2c); pink-circled (blue-circled) spot shows FWHM = 55 cm$^{-1}$ and $I_{2D}/I_G$ = 2.2 (FWHM = 43.8 cm$^{-1}$ and $I_{2D}/I_G$ = 2.91), indicating bilayer graphene. These results confirm that the CVD bilayer graphene film is highly homogeneous, with only very small fraction corresponding to possibly 3 layers.

We then quantified the bilayer graphene coverage by studying the statistics of 2D band peak width and $I_{2D}/I_G$ ratio. Figure 3d illustrates the histogram of the FWHMs of 2D band taken from the Raman mapping. The average peak width is determined to be $51 \pm 2$ cm$^{-1}$. Furthermore, cumulative counts plotted in Fig. 3e indicate that more than 99% of the FWHM values are below 60 cm$^{-1}$ (pink spheres). In addition, the histogram of $I_{2D}/I_G$ ratios (Fig. 3d, inset) shows an average value of $2.4 \pm 0.4$, and the corresponding cumulative count plot (Fig. 3e, inset) shows that more than 99% of $I_{2D}/I_G$ ratios are larger than 1. Using FWHM = 60 cm$^{-1}$ together with $I_{2D}/I_G = 1$ as the crossover values between bilayer and trilayer graphene, our data give an estimate of at least 99% coverage of bilayer graphene with less than 1% of trilayer over the entire area.

A direct verification of the bilayer nature of our CVD graphene film comes from electrical transport measurements. For this purpose, dual-gate bilayer graphene transistors were fabricated with three different dimensions, channel length and channel width of 1 μm×1μm, 1 μm×2μm, and 2 μm×2μm, respectively. A scanning electron microscope (SEM) image and an illustration of the fabricated device are shown in Fig. 4a. All devices have a local top gate and a universal silicon bottom gate with Al$_2$O$_3$ (40nm) and SiO$_2$ (310 nm) as the respective gate dielectrics. This dual-gate structure allows simultaneous manipulation of bilayer graphene bandgap and the carrier density by independently inducing electric fields in both directions.[15,16]

Figure 4b shows a two dimensional color plot of square resistance $R_\square$ vs. both top gate voltage ($V_{tg}$) and bottom gate voltage ($V_{bg}$), obtained from a typical 1×1 μm device at 6.5 K. The red and blue colors represent high and low square resistance, respectively. The data clearly show that $R_\square$ reach peak values along the diagonal (red color region), indicating a series of charge neutral points (Dirac points) when the top displacement fields cancel out the bottom displacement fields.[15,16] More importantly, the peak square resistance, $R_{\square, Dirac}$, reaches maximum at the upper left and lower right corner of the graph, where the average displacement fields from top and bottom gates are largest. Horizontal section views of the color plot in Fig. 4b are also shown in Fig. 4c, with $R_{\square}$ plotted against $V_{tg}$ at fixed $V_{bg}$ from -100 to 140 V. Once again, for each $R_{\square}$ vs. $V_{tg}$ curve square resistances exhibit a peak value, and $R_{\square, Dirac}$ increases with increasing $V_{bg}$ in both positive and negative direction. The charge neutral points are further identified in Fig. 4d in terms of the ($V_{tg}$, $V_{bg}$) values at $R_{\square, Dirac}$. Linear relation between $V_{tg}$ and $V_{bg}$ is observed with a slope of -0.073, which agrees with the expected value of $-\varepsilon_{bg}d_{tg}/\varepsilon_{tg}d_{bg}$ = -0.067, where $\varepsilon$ and $d$ correspond to the dielectric constant and thickness of the top gate (Al$_2$O$_3$: $d_{tg}$ = 40nm, $\varepsilon_{tg}$=7.5) and bottom gate (SiO$_2$: $d_{bg}$ = 310nm, $\varepsilon_{bg}$=3.9) oxide.[15,16] We also notice the deviation from linear relation at high field; the origin of which is not understood currently and requires further study.

Similar results from three other devices are shown in Supporting Information Fig. S3, and more than 46 measured devices show qualitative agreement. These electrical characterizations yield direct evidence for the successful synthesis of bilayer graphene. The observation of increasing $R_{\square, Dirac}$ values at higher fields is an unmistakable sign of bandgap opening in bilayer graphene.[15,16] In comparisons, the peak resistance at the charge neutral point should remain roughly constant for single-layer graphene,[15] while $R_{\square, Dirac}$ decreases at higher field for trilayer graphene.[5] In addition, we also compared the temperature dependence of $R_{\square, Dirac}$ at $V_{bg}$ ~ 0V and $V_{bg}$ ~ -100 V (Supporting Information Fig. S5). Larger variation of $R_{\square, Dirac}$ vs. temperature is observed under higher electric field, which again agrees with field-induced bandgap opening in bilayer graphene.[15,16] We note that the observed resistance modulation due to electric field and temperature are smaller compared to devices made by mechanical exfoliation,[15,16] which can be attributed to the polycrystalline nature of CVD graphene film. We also note that our devices show large fluctuations of the offset voltage (from impurity and surface doping), with some cases exceeding 140 V for the bottom gate. This could be caused by the ion residues from the etching process, and further investigations are needed.

We also studied the statistics of bilayer graphene occurrence for 63 (7 row x 9 columns) dual-gate devices fabricated across the same film (Fig. 5a). 46 out of 63 devices show bilayer graphene behaviors, characterized by increasing $R_{\square,Dirac}$ at larger fields. Of the remaining devices, 2 devices contain no graphene pieces, and 14 have fabrication defects.[34] Interestingly, one device shows trilayer characteristics[5] with decreasing $R_{\square,Dirac}$ under both more positive and more negative fields (Supporting information Fig. S4). Hence, 46 out of 47 (98%) working devices show bilayer characteristics. For the bilayer graphene devices, we also calculated the maximum percentage changes of peak square resistance, $\Delta R_{\square,Dirac} / R_{\square,Dirac,min}$, in which $\Delta R_{\square,Dirac}$ denotes the maximum difference in $R_{\square,Dirac}$ within $V_{tg}$ of $\pm10V$ and $V_{bg}$ of $\pm120V$, and $R_{\square,dirac,min}$ is the minimum peak square resistance. The histogram of the percentile changes is shown in Fig. 5b, with an average peak resistance change of 38% and maximum value of 77%. In addition, the average room temperature carrier mobilities were measured to be ~580 cm$^2$V$^{-1}$s$^{-1}$, which are the lower-bound values without excluding the device contact resistance. The smaller-than-expected $R_{\square,Dirac}$ modulation is believed to be caused by defects and unintended impurity doping[16]. High quality gate dielectrics have been shown to improve the bilayer graphene device performance dramatically[18]. The electrical measurement results echo the finding from Raman measurements: our CVD grown bilayer graphene film is highly homogeneous.

Lastly, we would like to comment on the key growth parameters for our CVD bilayer graphene films. It has been suggested that graphene growth on copper surface is self-limited to single layer[24], but both of our Raman and electrical characterizations clearly prove otherwise. We systematically varied the key growth conditions, and the resulting film quality was evaluated using Raman spectroscopy (Supporting Information, and Table S1). In brief, increasing CH$_4$ flow rate by 2 times has no noticeable effect on 2D band width and $I_{2D}/I_G$ values, except for a larger $I_D/I_G$ ratio corresponding to more disorders. However, increasing growth pressure to ambient condition leads to larger 2D band width and smaller $I_{2D}/I_G$ ratio, indicating the increasing portion of trilayer graphene. This result is consistent with recent literature, that higher pressure favors multilayer graphene growth on copper surface.[35] Based on our results, we speculate that the key parameter for the bilayer graphene film growth is the slow cooling process (~18°C/min). Cooling rate has been found to be the critical factor for forming uniform single or bilayer graphene on Nickel.[21,36,37] Our initial results should promote studies of the detailed growth mechanism for bilayer graphene.

The size of the homogeneous bilayer graphene films is limited only by the synthesis apparatus, which can be further scaled up. The integration with existing top-down lithography techniques should bring significant advancement for high performance, light-weight, and transparent graphene electronics and photonics. Furthermore, because the CVD grown bilayer graphene film can be transferred to arbitrary substrates, adopting high-k dielectrics for both top and bottom gates should drastically improve the device performance.[18] A few voltages applied to the gate electrodes will be able to open up sizeable bandgap (~250 meV).

**Acknowledgment.** We thank Prof. A. Matzger for his assistance in Raman mapping, and Dr. Kai Sun for his assistance in TEM charaterization. We also thank Chang-Hua Liu and Rui Li for early assistance in sample preparation. The work is supported by the startup fund provide by the University of Michigan. This work used the Lurie Nanofabrication Facility at University of Michigan, a member of the National Nanotechnology Infrastructure Network funded by the National Science Foundation.

**Supporting Information Available.** Growth & transfer process, device fabrication, the table of key growth parameters, figures for Lorentzian fitting of Raman spectrum, additional 2-dimensional color plot of bilayer electrical transport and trilayer electrical transport characteristics, and temperature-dependent resistance change. This material is available free of charge via the Internet at http://pubs.acs.org.

**FIGURE CAPTIONS**

FIGURE 1. Wafer scale homogeneous bilayer graphene film grown by CVD. (a) Photograph of a 2 inch × 2 inch bilayer graphene film transferred onto a 4 inch Si substrate with 280nm thermal oxide. (b) Optical microscopy image showing the edge of bilayer graphene film. (c) AFM image of patterned bilayer graphene transferred onto $SiO_2$/Si substrate. (Inset) Cross section height profile measured along the dotted line. (d) Raman spectra taken from CVD grown bilayer graphene (red solid line), exfoliated single-layer (green solid line) and bilayer graphene (blue solid line) samples. Laser excitation wavelength is 514 nm.

FIGURE 2. Selected area electron diffraction pattern of bilayer graphene. (a) Normal incident diffraction pattern of bilayer graphene sample. The bilayer graphene film was transferred onto copper grid with holy carbon supporting film. The diffraction image was taken by JEOL 2010F Analytical Electron Microscope with acceleration voltage of 200 kV. (b) Profile plot of diffraction peak intensities across a line cut indicated by the green arrows shown in (a). (c) Diffraction peak intensities as a function of tilt angle for (-1-120) (in red) and (0-110) (in blue).

FIGURE 3. Spatially resolved Raman spectroscopy of CVD bilayer graphene. (a) and (b), Two-dimensional color mapping of the FWHMs of Raman 2D band and $I_{2D}/I_G$ ratios over 30 μm × 30 μm area, respectively. (c) Raman spectra from the marked spots corresponding colored circles showing bilayer and trilayer graphene. (d) Histogram of the FWHMs of Raman 2D band corresponding to area shown in (a). (Top right Inset) Histogram of $I_{2D}/I_G$ ratios for the same area. (e) Cumulative count plot of FWHMs of 2D band. Pink (blue) spheres represent the FWHM less (more) than 60 cm$^{-1}$. (Inset) Cumulative count plot of $I_{2D}/I_G$ ratios. Pink (blue) spheres indicate the ratio larger (smaller) than 1. (For Raman mapping, $\lambda_{laser}$=514 nm, 500nm step size, 100x objector).

FIGURE 4. Electrical transport studies on dual-gate bilayer graphene devices. (a) Scanning electron microscopy image (top) and illustration (bottom) of a dual-gate bilayer device. The dashed square in the SEM image indicates the 1μm × 1μm bilayer graphene piece underneath the top gate. (b) Two dimensional color plot of square resistance $R_\square$ vs. top gate voltage $V_{tg}$ and back gate voltage $V_{bg}$ at temperature of 6.5K. (c) $R_\square$ vs. $V_{tg}$ at different value of fixed $V_{bg}$. The series of curves are taken from $V_{bg}$ of -100V to 140V, with 20V increment. (d) The charge neutral points indicated as set of ($V_{tg}$, $V_{bg}$) values at the peak square resistance $R_{\square,dirac}$. The red line is the linear fit. The electrical measurements were carried out in a closed cycle cryogenic probe station (LakeShore, CRX-4K), using lock-in technique at 1kHz with AC excitation voltage of 100μV.

FIGURE 5. Bilayer statistics from electrical transport measurement on dual-gate graphene devices. (a) A color-coded map of 63 devices (7 rows x 9 columns) fabricated across the same graphene film. The red squares indicate bilayer graphene confirmed by transport measurement; the yellow squares indicate devices which have fabrication defects; the white squares mark the region with no graphene; and the green square represents device with trilayer response from the transport measurement. (b) Histogram of $\Delta R_{\square,dirac}$ / $R_{\square,dirac,min}$ values in percentage for 46 active devices. $\Delta R_{\square,dirac}$ corresponds to the maximum difference in $R_{\square,dirac}$ within $V_{tg}$ of ±10V and $V_{bg}$ of ±120V. $R_{\square,dirac,min}$ is the minimum peak resistance.

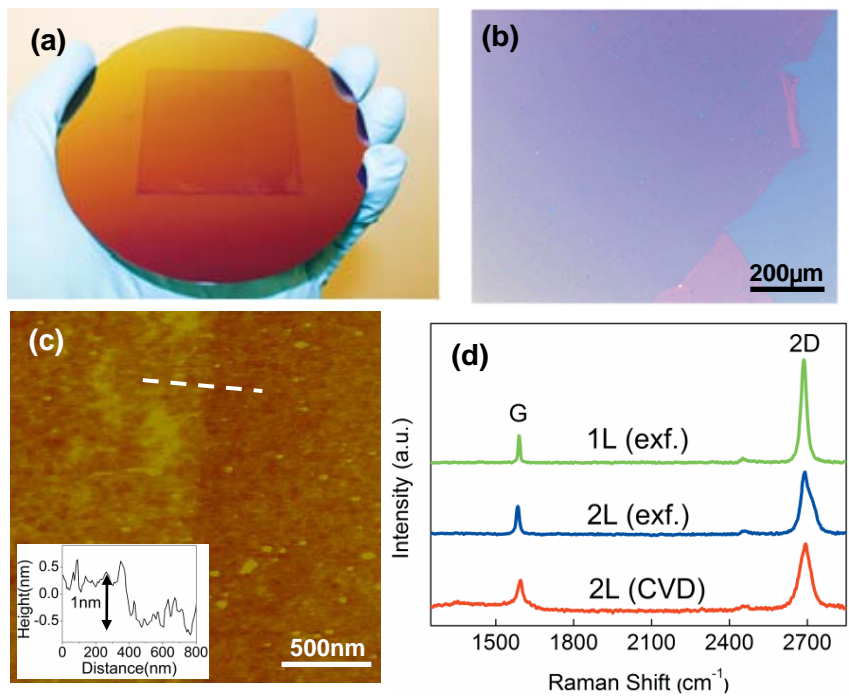

**FIGURE 1.**

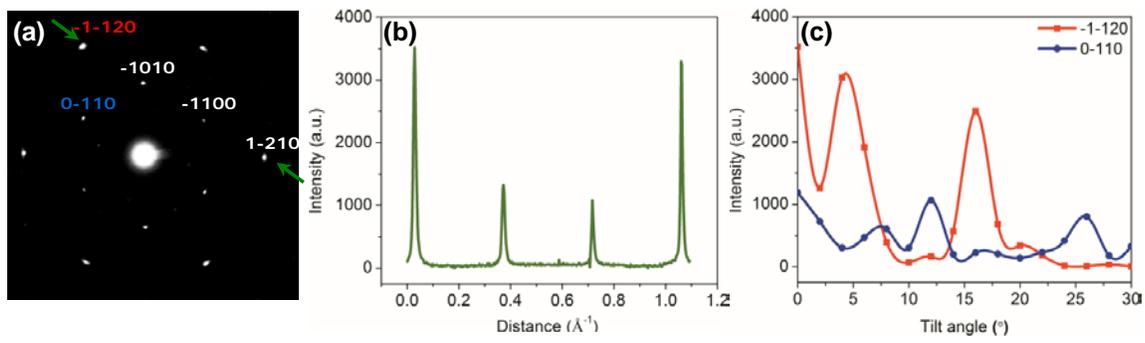

**FIGURE 2.**

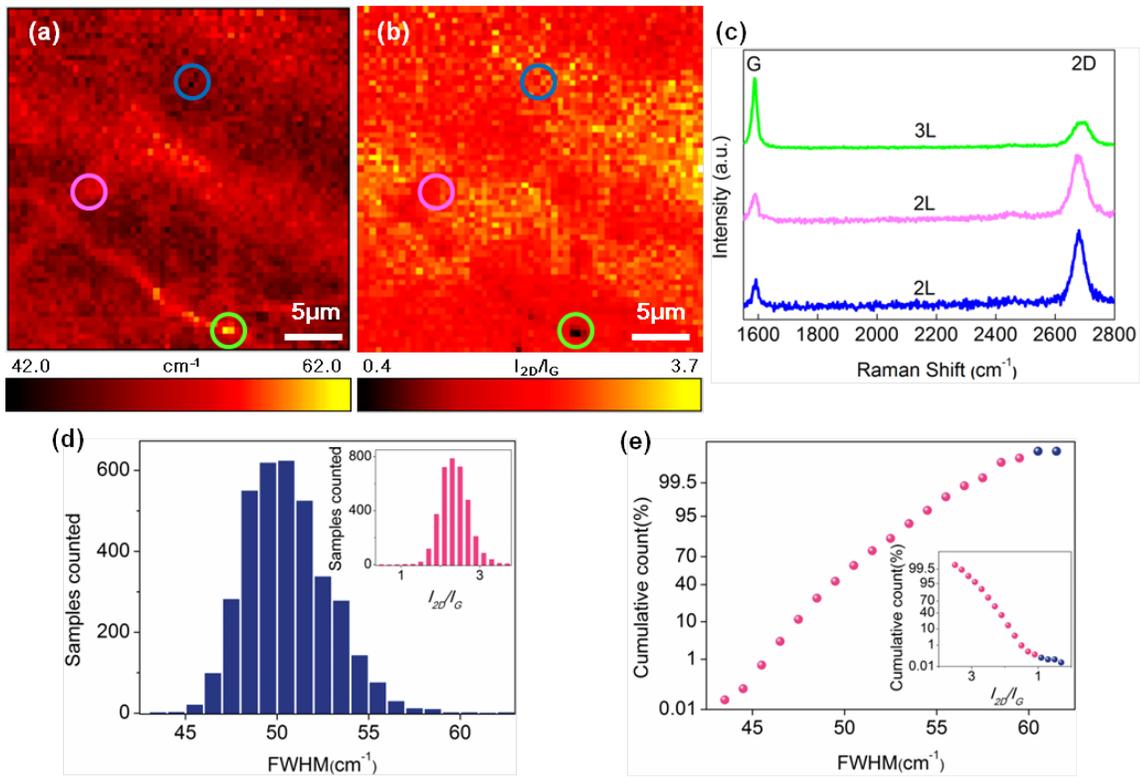

**FIGURE 3.**

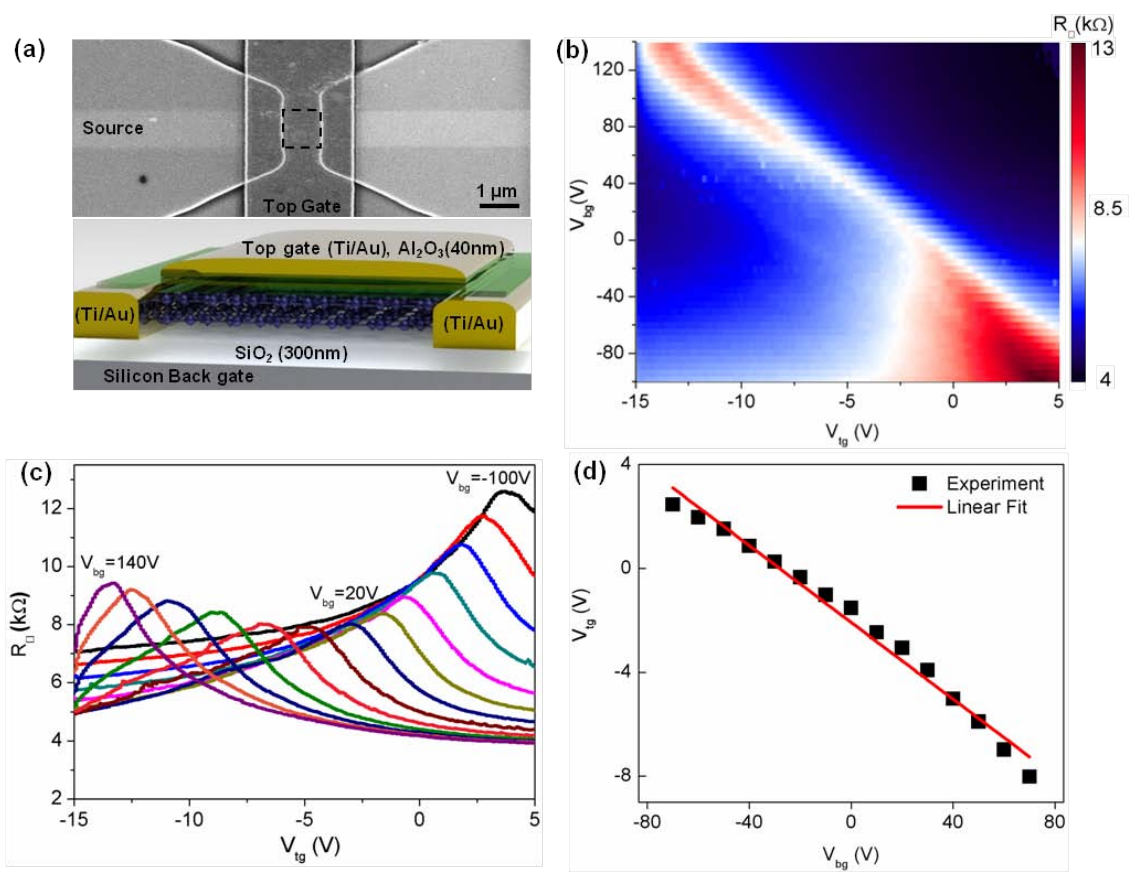

**FIGURE 4.**

**(a)**

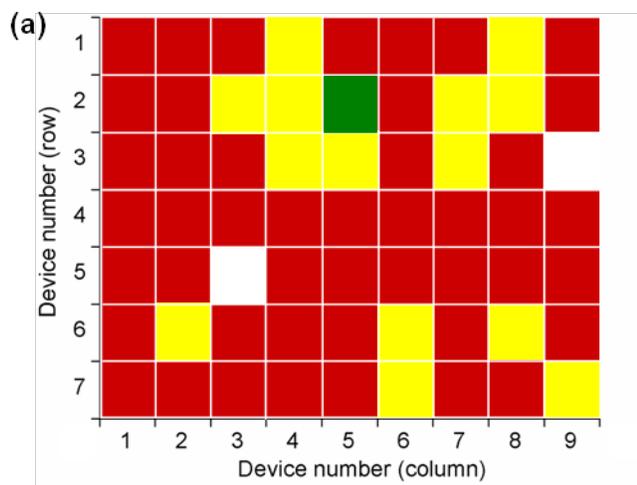

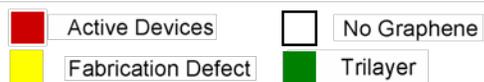

**(b)**

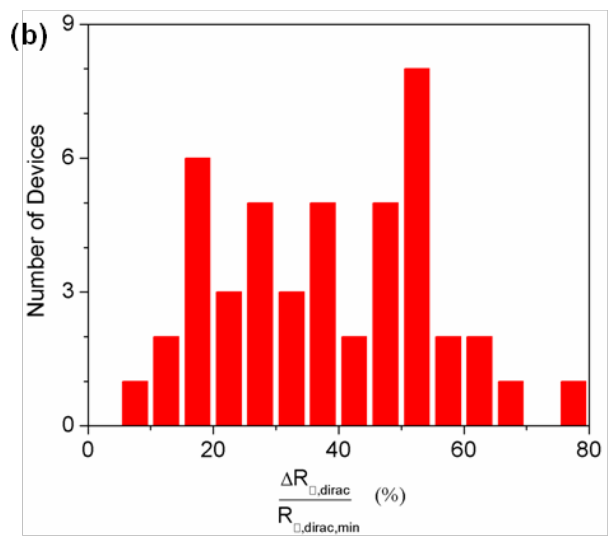

**FIGURE 5.**

**Supporting Information for**

# Wafer Scale Homogeneous Bilayer Graphene Films by

# Chemical Vapor Deposition

Seunghyun Lee, Kyunghoon Lee, Zhaohui Zhong

Department of Electrical Engineering and Computer Science, University of Michigan, Ann Arbor, MI 48109

**Bilayer graphene growth & transfer**

25µm thick copper foil (99.8%, Alfa Aesar) was loaded into an inner quartz tube inside a 3 inch horizontal tube furnace of a commercial CVD system (First Nano EasyTube 3000). The system was purged with argon gas and evacuated to a vacuum of 0.1 Torr. The sample was then heated to 1000°C in $H_2$ (100 sccm) environment with vacuum level of 0.35 Torr. When 1000°C is reached, 70 sccm of $CH_4$ is flowed for 15 minutes at vacuum level of 0.45 Torr. The sample is then cooled slowly to room temperature with a feed back loop to control the cooling rate. The vacuum level is maintained at 0.5 Torr with 100 sccm of argon flowing. The time plot of the entire growth process is shown in Fig. S1.

Two different methods were used to transfer bilayer graphene from copper foil to $SiO_2$ substrates. The first method utilize thermal release tape (Nitto Denko) to transfer bilayer graphene from the copper foil.[1] The tape was attached to the copper surface and a force of 6.25 $N/cm^2$ was applied to the copper/bilayer graphene/tape stack for 10 minutes with EVG EV520IS wafer bonder. The other side of the copper is exposed to $O_2$ plasma for 30 seconds to remove the graphene on that side. The copper was etched away using iron (III) nitrate (Sigma Aldrich) solution (0.05g/ml) for 12 hours. A 4 inch silicon wafer with thermally grown $SiO_2$ was precleaned with nP12 nanoPREP using plasma power of 500W for 40 seconds to modify the surface energy and produce a hydrophilic surface. The tape and bilayer graphene stack was transferred to the precleaned $SiO_2$ wafer and a force of $12.5N/cm^2$ was applied for 10 minutes. The substrate was then heated to 120 °C to eliminate the adhesion strength of the thermal release tape. The tape was then peeled off and the adhesive residue was removed with warm acetone.

Polymethyl methacrylate (PMMA) can also be used instead of thermal release tape to transfer bilayer graphene.[2] This method is easier as it does not require a bonding tool but the edge part of the graphene is usually rough due to uneven thickness of spin coated PMMA at the edge. In this method, one side of the sample is coated with 950PMMA A6 (Microchem) resist and cured at 180°C for 5 minutes. The other side of the sample is exposed to $O_2$ plasma for 30 seconds to remove the graphene on that side. The sample is then left in iron (III) nitrate (Sigma Aldrich) solution (0.05g/ml) for at least 12 hours to completely dissolve away the copper layer. The sample is transferred on to a silicon substrate with thermal oxide. The PMMA coating is removed with acetone and the substrate is rinsed several times. After all transfers, Raman spectroscopy as well as optical microscope were used to characterize the graphene film. Electrical transport measurement was done with samples prepared with PMMA method.

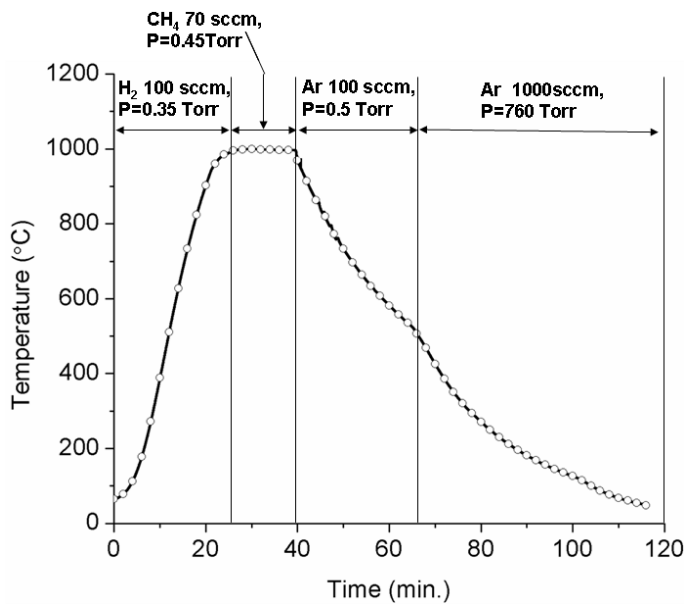

**Figure S1.** Temperature vs. time plot of bilayer graphene growth condition. Pressure value is denoted as "P".

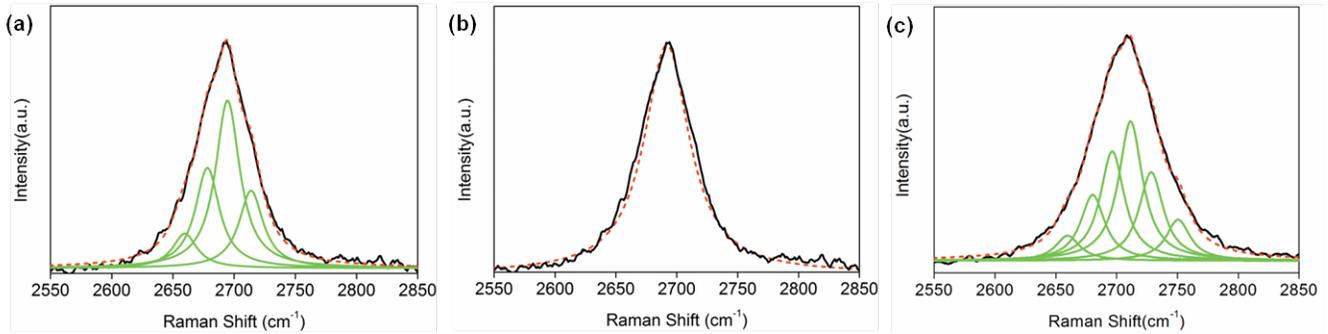

**Figure S2.** (a) The measured 2D Raman band of a bilayer with the FWHM of 45cm$^{-1}$. The peak can be well-fitted with the sum of four single Lorentzian (green solid line) of 24cm$^{-1}$ FWHM. (b) Single Lorentzian fit (red dash line) of the same data in Fig. S2a clearly shows deviation from the measured 2D band. (c), The measured 2D Raman band of a trilayer with the FWHM of 62cm$^{-1}$. 2D peak of trilayer are fitted with six single Lorentzian (green solid line)

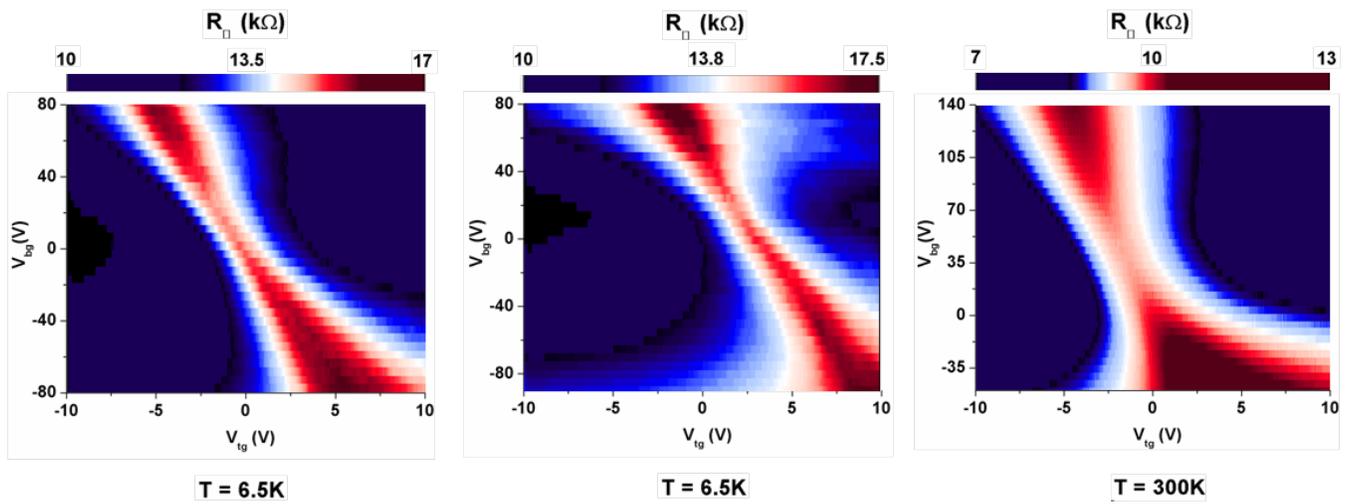

**Figure S3.** Three dual-gate graphene devices showing bilayer transport behaviour.

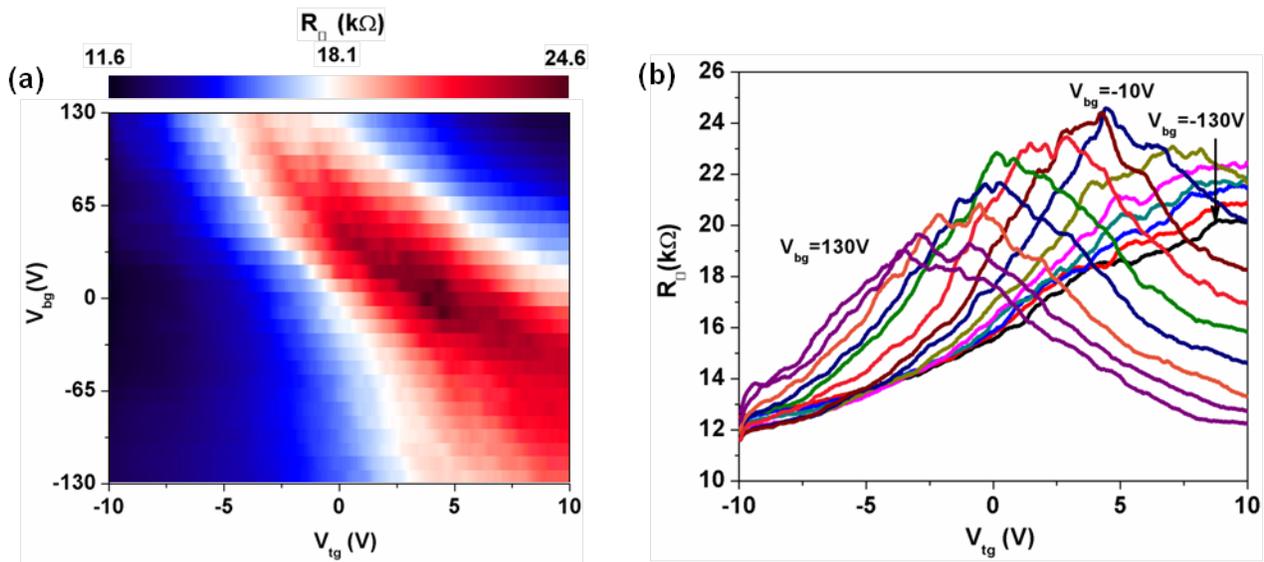

**Figure S4.** (a) A device showing trilayer transport behaviour. The observed peak square resistance decreases as increasing field. This is distinctively different from bilayer response. (b) Horizontal section views with $R_\square$ plotted against $V_{tg}$ at fixed $V_{bg}$ from -130 to 130 V with 20V increment.

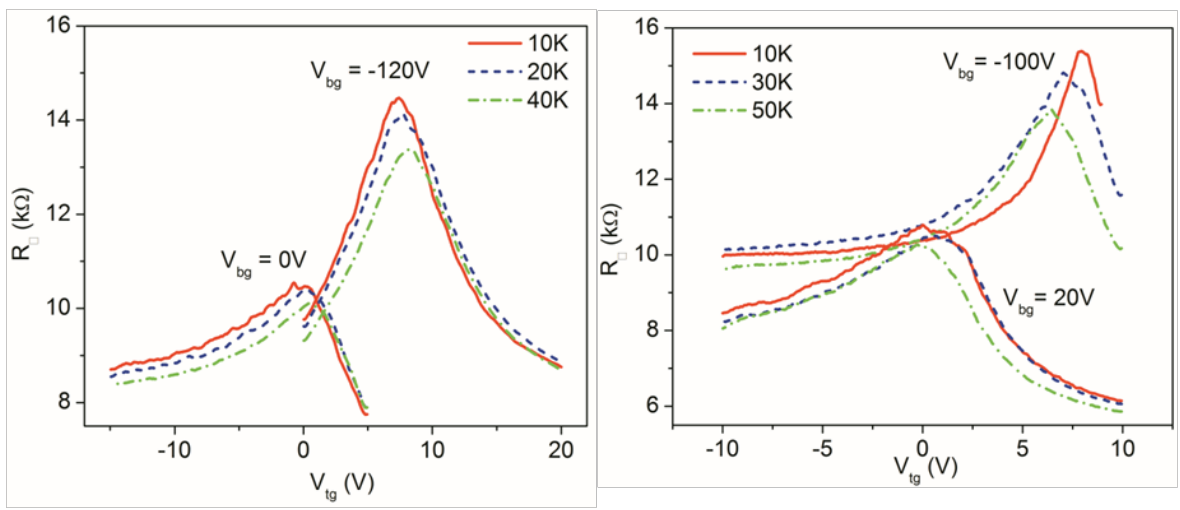

**Figure S5.** Two dual-gate graphene devices showing temperature dependent resistance versus top gate voltage sweep at two different back gate voltages.

| Sample No. | Growth Pressure (Torr) | Growth Temperature (°C) | Growth Time (min) | Ar Flow rate (sccm) | $CH_4$ Flow rate (sccm) | $H_2$ Flow rate (sccm) | 2D Band FWHM ($cm^{-1}$) | $I_{2D}/I_G$ | $I_D/I_G$ | Cooling rate (°C/min) |
|---|---|---|---|---|---|---|---|---|---|---|
| 1 | 0.5 | 1000 | 15 | 0 | 70 | 0 | 46.6 | 2.628 | 0.258 | 18 |
| 2 | 0.5 | 1000 | 15 | 0 | 140 | 0 | 47 | 2.12 | 0.57 | 18 |
| 3 | Ambient | 1000 | 15 | 1000 | 50 | 0 | 59.12 | 1.402 | 0.36 | 18 |
| 4 | 1.5 | 1000 | 15 | 0 | 40 | 600 | 60 | 0.64 | 1.11 | 18 |

**Table S1.** Comparison of graphene samples synthesized under different conditions.